\documentclass[sn-basic]{sn-jnl}
\theoremstyle{thmstyleone}%
\theoremstyle{thmstyletwo}%

\theoremstyle{thmstylethree}%

\begin{document}

\title[Article Title]{A comparison of cosmological models with high-redshift quasars}


\author[1]{\fnm{Liuyuan} \sur{Fan}}

\author[1]{\fnm{Guanwen} \sur{Fang}}

\author*[2]{\fnm{Jian} \sur{Hu}}
\email{ dg1626002@smail.nju.edu.cn}

\affil[1]{\orgdiv{Institute of Astronomy and Astrophysics}, \orgname{Anqing Normal University}, 
\city{Anqing}, \postcode{246133}, 
\country{People’s Republic of China}}

\affil*[2]{\orgdiv{Department of Engineering}, \orgname{Dali University}, 
\city{Dali }, \postcode{671003}, 
\country{People’s Republic of China}}



\abstract{The non-linear relationship between the monochromatic X-ray and UV luminosities in quasars offers the possibility of using high-z quasars as standard candles for cosmological testing. In this paper, we use a high-quality catalog of 1598 quasars extending to redshift $\sim 6$, to compare the flat and uniformly expanding cosmological model,\textbf{ $R_{\rm h}=ct$ and $\Lambda$CDM cosmological models} which are the most debated. The quasar samples are mainly from the XMM-Newton and the Sloan Digital Sky Survey (SDSS). \textbf{The final result is that} the Akaike Information Criterion favors $\Lambda$CDM over $R_{\rm h}=ct$ with a relative probability of $\sim 86.30\%$ versus $\sim 13.70\%$.}



\maketitle

\section{Introduction}\label{sec1}
Although the cosmological constant($\Lambda$) is supposed to be the simplest and best candidate for dark energy and accounts for many remarkable cosmological observations (\citealp{1998AJ....116.1009R, 1999ApJ...517..565P, 2011ApJS..192...17B, 2013ApJS..208...20B, 2014A&A...571A..16P, 2016A&A...594A..13P, 2005ApJ...633..560E, 2015NewAR..67....1W}), it also needs to face some flaws such as the fine-tuning and cosmic coincidence problems (\citealp{1989RvMP...61....1W, 1999PhRvL..82..896Z}). Meanwhile, the so-called power-lower cosmology with the assumption that the scale factor evolves as a(t) $\propto t^{n}$ (\citealp{1997PhRvD..55.5881D, 2014JCAP...10..047D}) is notable. \textbf{Moreover}, as a particular case of the power-law cosmology with an exponent equal to 1, a promising alternative of the standard model, $R_{\rm h}=ct$ cosmology (\citealp{2007MNRAS.382.1917M, 2012MNRAS.419.2579M}) explains a mass of physics well, such as the epoch of reionization (\citealp*{2016MNRAS.456.3422M}), the cosmic microwave background (CMB) multiple alignments (\citealp*{2015AJ....149....6M}), and so on. A study combined the \textbf{quasar-stellar} object Hubble diagram (QSO HD) with the Alcock-Paczy'nski test, which is entirely independent of any galaxy evolution, carried out a comparative analysis of nine different cosmological models (\citealp*{2016IJMPD..2550060L}), however, only $\Lambda$CDM/wCDM and $R_{\rm h}=ct$ passed the combined tests. While other models, such as the Milne universe, Einstein-de Sitter, and the Static universe\textbf{,} were excluded at the $>99\%$ confidence level and so forth. Comparisons between the two cosmological models have continued over the past decades, using the latest data and methods.

\textbf{For example, \cite{2016PhRvD..94j3511T} derived that $\Lambda$CDM model was statistically superior to the power-law cosmological model by simultaneously combining three standard probes, namely type Ia supernovae (SNIa), baryonic acoustic oscillations (BAO), and acoustic scale information from the cosmic microwave background (CMB), using the model selection tools Akaike (1973) Information Criterion (AIC) and Bayesian Information Criterion (BIC) (\citealp*{1978AnSta...6..461S}) which will be compared in the chapter of methods.}

Another approach was to use the Markov Chain Monte Carlo (MCMC) method to constrain the cosmological parameters and quantitatively \textbf{decided} the most appropriate model through Bayes factors (\citealp*{2019MNRAS.484.4337T}), based on the 152 strong gravitational lensing systems, 30 H(z), and 11 BAO data. The additional data, namely, the 30 H(z) and 11 BAO data, were added to compensate for the lack of capacity of strong gravitational lensing systems to constrain cosmological parameters, which is a straightforward and effective way. The results also strongly \textbf{preferred} to support the Flat-$\Lambda$CDM and curve-$\Lambda$CDM rather than power-law and $R_{\rm h}=ct$ universe. Similar results could be obtained from cosmic distance-duality (CDD) relations tests (\citealp*{2018MNRAS.477.5064H}) as well.

\textbf{\cite{2012MNRAS.425.1664B} concluded that $R_{\rm h}=ct$ could be conclusively eliminated by using H(z) measurements from chronometers and BAO. However, \cite{2013MNRAS.432.2669M} argued that $R_{\rm h}=ct$ was favored over $\Lambda$CDM by using a similar dataset which only excluded BAO dataset. To resolve the contradiction between the two groups of authors, \cite{2020EPJC...80..694S} carried out $\Lambda$CDM  was decisively favored over $R_{\rm h}=ct$ for the unbinned/GPR reconstructed dataset by using the priors centered around the Plank 2018 best-fit $\Lambda$CDM values.}

However, there are also growing opposite opinions on whether the standard cosmology is better than $R_{\rm h}=ct$. \textbf{Plenty of counterviews have been published based on various integrated observational signatures, such as the angular size of galaxy clusters, the GRB Hubble diagram, and so forth. Table 2 of \cite{2018MNRAS.481.4855M} gives a more detailed summary of these analyses and results.} The results listed in this table almost all favor the $R_{\rm h}=ct$ universe. For instance, using the most up-to-date samples of 408 flat spectrum radio quasars (FSRQs) detected by the Fermi Large Area Telescope over its four-year survey, \cite{2016MNRAS.462.3094Z} demonstrated that the aforementioned gramma-emitting FSRQs are large enough to carry out meaningful cosmological testing between the $R_{\rm h}=ct$ and $\Lambda$CDM cosmologies. Subsequently, they found that these data implicitly support $R_{\rm h}=ct$ more than $\Lambda$CDM  based on the FSRQ gamma-ray luminosity function, even without optimization of $H_{0}$ for $R_{\rm h}=ct$.

In recent years, Hubble diagrams of high-redshift quasars have been developed based on various techniques. Nevertheless, based on the current Hubble diagram of high-redshift quasars, the one-to-one comparison between the $R_{\rm h}=ct$ universe and the standard model $\Lambda$CDM shows that the data of high-redshift quasars seem to be more inclined to support the former. In this paper, we will make a comparison between the two cosmological models once again. \textbf{Few previous data }include an adequately large redshift interval and, worse, some data dependent on the assumed background model\textbf{, which} is the principal reason why using the model-independent approach and the unprecedently large range of redshifts of the data. What we represent in this paper significantly pushed this issue forward through our analysis. 

The main structure of the paper is as follows. In section II, we will introduce the two cosmological models and describe the method adopted to obtain the result that which model is better adjusted to the data. In section III, we will discuss the results of the model comparison and make a conclusion in section IV.

\section{Methods}\label{sec11}

Based on the 7237 parent samples obtained from the cross-correlation of the third XMM-Newton serendipitous source catalog, data release 7 (3XMM-DR7) (\citealp*{2016A&A...590A...1R}), with the Sloan Digital Sky Survey (SDSS) quasar catalogs from data releases 7 (SDSS-DR7) (\citealp*{2011ApJS..194...45S}) and 12 (SDSS-DR12) (\citealp*{2017A&A...597A..79P}), \cite{2019NatAs...3..272R} produced final 1598 sources with reliable measurements of the intrinsic X-rays emission and the UV emission, avoiding any possible physical and observational contaminants that could bias these measurements. These data are interpreted according to a universally accepted framework; the UV emission comes from an accretion disk, where the gravitational energy of the accreting gas is converted into radiation (\citealp{1973A&A....24..337S}). \cite{1993ApJ...413..507H} found that the X-ray emission comes from the inverse Compton scattering of ultraviolet photons and the nonlinear $L_{X}$-$L_{UV}$ relations hold only for quasars that are blue in the UV and soft in the X-ray part of the spectrum. So in this paper, we will use the sources mentioned above and the relation to analyze the degree of matching between the models and the data.

When $L_{X}$ and $L_{UV}$ are the monochromatic luminosities at rest frames observed at 2 Kev and 2500 $\dot{A}$ respectively, the non-linear relationship between the UV and X-ray emissions of quasars is usually expressed as 
\begin{equation}
	{\log _{10}}{L_X} = \gamma {\log _{10}}{L_{UV}} + \beta. 
\end{equation}
The key point of the application for cosmological testing of this equation is the non-evolution of the relation with the redshift between ${L_X}$ and ${L_{UV}}$. Some authors have also found that the slope of this relationship ${\gamma}$ varies in a small range, ${\gamma  \sim 0.5{\rm{ - 0}}{\rm{.7}}}$ (\citealp*{2003AJ....125..433V, 2006AJ....131.2826S, 2007ApJ...665.1004J}). To get the utmost out of the database, based on the relationship between the fluxes and absolute luminosities, we improve equation~(1) to the following form
\begin{equation}
	{\log _{10}}{F_X} = \tilde \beta  + \gamma {\log _{10}}{F_{UV}} + 2(\gamma  - 1){\log _{10}}{d_L},
\end{equation}
where $\tilde\beta$ is a constant that embodies the slope $\gamma$ and the intercept $\beta$ in equation~(1), that is,
\begin{equation}
	\tilde \beta  = \beta  + (\gamma  - 1){\log _{10}}4\pi.
\end{equation}
Here, by maximizing the likelihood function
\begin{equation}
	\ln (LF) =  - \frac{1}{2}\sum\limits_{i = 1}^{1598} {\left\{ {\frac{{{{\left[ {{{\log }_{10}}{{({F_X})}_i} - \Phi ({{[{F_{UV}}]}_i},{d_L}[{z_i}])} \right]}^2}}}{{\tilde \sigma _i^2}} + \ln (2\pi \tilde \sigma _i^2)} \right\}} ,
\end{equation}
where the variance $\tilde \sigma _i^2 \equiv {\delta ^2} + \sigma _i^2$ is given based on the global intrinsic dispersion, $\delta$, and the measurement error $\sigma_{i}$ in $(F_{X})_{i}$ (\citealp*{2015ApJ...815...33R}), we optimized each models' parameters along with those characterizing the empirical fit in equation~(2). As for all the sources in the sample, the rest-frame UV data provide higher quality flux measurements than the X-ray ones. And compared to the $\sigma_{i}$ and $\delta$, the error in ${({F_{UV}})_i}$ is so small that it can be ignored. The function $\Phi$ in equation~(4) is defined as
\begin{equation}
	\Phi ({[{F_{UV}}]_i},{d_L}[{z_i}]) = \tilde \beta  + \gamma {\log _{10}}{({F_{UV}})_i} + (2\gamma  - 2){\log _{10}}{d_L}({z_i}),
\end{equation}
in accordance with the measured fluxes $(F_{UV})_{i}$ and $(F_{X})_{i}$ at redshift $z_{i}$.

The luminosity distance in these models are 
\begin{equation}
	d_L^{{R_h} = ct}(z) = \frac{c}{{{H_0}}}(1 + z)\ln (1 + z); \label{num6}
\end{equation}
\begin{equation}
	d_L^{\Lambda CDM}(z) = \frac{c}{{{H_0}}}(1 + z)\int_0^z {\frac{{du}}{{\sqrt {{\Omega _m}{{(1 + u)}^3} + {\Omega _\Lambda }} }}}  
\end{equation}
respectively. The Hubble constant $H_{0}$ in the two models is not independent of $\tilde \beta$, and it will be optimized along with this parameter. In this paper, we \textbf{adopt} the $70 km{s^{ - 1}}Mp{c^{ - 1}}$ as the Hubble parameter, if one prefers a different value $H'_0$ then, the optimized value of $\tilde \beta$ just needs to be changed by an amount $\Delta \tilde \beta  = (2\gamma  - 2){\log _{10}}({{{{H'}_0}} \mathord{\left/
		{\vphantom {{{H'_0}} {{H_0}}}} \right.\kern-\nulldelimiterspace} {{H_0}}})$, according to Equation~(2). It should be noted that the number of parameters in the two models is different, which will have a crucial influence in judging the model fits by using an information criterion. $R_{\rm h}=ct$ universe has no parameters due to the previously assumed Hubble parameter, $\Lambda$CDM has one. Since the $L_{X}$-$L_{UV}$ relation of the quassar is used in the fitting process, the number of their parameters is 3 and 4, respectively.

It is obvious that more parameters could improve the fitting degree of the model to the dataset, so the comparison of the maximum likelihood function of different models always supports the model with more parameters, although it may induce over-fitting. The information criterion avoids the problem of over-fitting by introducing the penalty term of model complexity to find the best balance and the model’s ability to describe data. The penalty term could avoid the phenomenon that with more parameters, the cosmology model may adjust the noise as well. Finally, the information criterion describes the relative evidence for one model over another through relative likelihood. The smaller the value of the information criterion, the better the model. Since the parameters in the two models are different and the process of the data as we shall see, a simple $\chi^{2}$ minimization or a comparison of the likelihood function is inadequate to judge which cosmology is better.

From equation~(4), we get the maximum likelihood function, and the AIC is defined by $AIC = 2k-2ln(LF)$, where k is the number of free parameters in each model. \cite{2019MNRAS.489..517M} uses BIC, which is defined by ${BIC{\rm{ = }}k\ln (n) - 2\ln (LF)}$, where ${n}$ is the number of the sample, to compare different models. Both of these information criteria are used to determine the goodness of the model. The penalty term inside the AIC is ${2k}$. The penalty term in the BIC is not only related to the number of parameters of the model, but also to the number of samples, which can be written as ${k\ln(n)}$. \textbf{We use a sample of the quasar,} which has a large deviation and poor quality compared to a standard candle, SNe Ia, for the same number of cases; therefore, we adopt a penalty term independent of the sample size and only related to the model parameters to determine the model, and here AIC is chosen. The relative likelihood of model $\alpha$ being correct is $P(\alpha ) = \frac{{\exp ({{ - AI{C_\alpha }} \mathord{\left/
				{\vphantom {{ - AI{C_\alpha }} 2}} \right.\kern-\nulldelimiterspace} 2})}}{{\sum\nolimits_\beta  {\exp ({{ - AI{C_\beta }} \mathord{\left/{\vphantom {{ - AI{C_\beta }} 2}} \right.\kern-\nulldelimiterspace} 2})} }}$, where $ \exp ({{ - AI{C_\alpha }} \mathord{\left/{\vphantom {{ - AI{C_\alpha }} 2}} \right.\kern-\nulldelimiterspace} 2})$ is its Akaike weight. 
			
\section{Results}\label{sec2}

In the paper of \cite{2019NatAs...3..272R}, they divided the quasar samples into subsamples within confined redshift bins that were smaller than the measured dispersion. Then they \textbf{fitted} the log-linear relation ${\log _{10}}{L_X} = \gamma {\log _{10}}{L_{UV}} + \beta$ in each narrow redshift bin to calculate the finally average value of $\gamma$. And used the SNe Ia as external calibrators to estimate the $\beta$ by cross-matching the Joint Light-curve Analysis type Ia supernovae sample with the quasar sample in the overlapping redshift range. But here, since the internal self-calibration can do better than the external calibration when approving large samples consisting of different sub-samples (\citealp*{2015AJ....149..102W, 2018A&A...610A..87M}), the method we adopt is using the quasar Hubble diagram to optimize all the parameters including those of the model itself and the $\gamma$, $\beta$ or $\delta$ appearing in Equations~(1)-(5) simultaneously. Namely, the minimum $\chi^2$-fitting of the data with the model is performed to calibrate the models' parameters inside. As we will see next, the values of $\gamma$ and $\delta$ change slightly, but $\Omega_{M}$ changes significantly, compared to the results \cite{2019NatAs...3..272R} got by using the external calibrators.

\begin{table*}
\tabcolsep=0.1cm
\begin{center}
\caption{ Results of model comparison based on high-redshift quasars}
\begin{tabular}{lcccccc}
&&&&&& \\
\hline\hline
&&&&&& \\
Model& ${\beta}$ & $\gamma$ & $\delta$ & $\Omega_{\rm m}$ & AIC & Likelihood  \\
&&&&&& \\
\hline
&&&&&& \\  
$\Lambda$CDM & ${\rm{ - }}11.38_{{\rm{ - }}0.24}^{{\rm{ + }}0.25}$ & $0.622\pm0.01$ & $0.231\pm0.004$ & $0.622_{{\rm{ - }}0.17}^{{\rm{ + }}0.19}$ & -42.01 & $86.30\%$ \\
&&&&&& \\
$R_{\rm h}=ct$ & ${\rm{ - }}11.01_{{\rm{ - }}0.18}^{{\rm{ + }}0.17}$ & $0.639_{{\rm{ - }}0.009}^{{\rm{ + }}0.008}$ & $0.231\pm0.004$ & --- & -38.33 & $13.70\%$ \\
&&&&&& \\
\hline\hline
\end{tabular}
\end{center}
\end{table*}

\begin{figure}
	\centering
	\includegraphics[scale = 0.43]{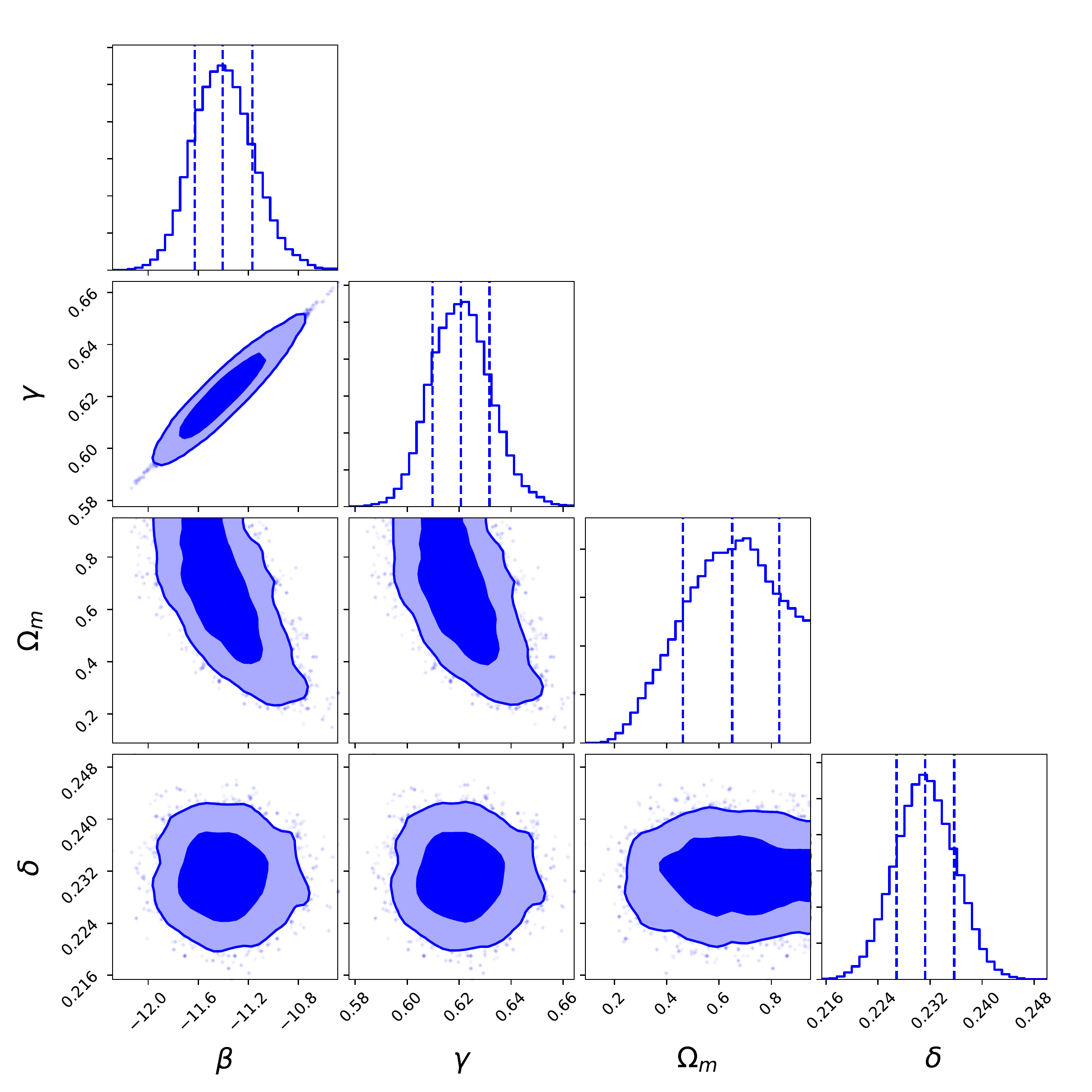}
	\label{fig1}
	\caption{Fitting results of $\Lambda$CDM models from high-z quasar samples.The two areas represent $1\sigma$ and $2\sigma$ confidence intervals respectively. The optimized values of these correlational parameters for this model are $\gamma \approx 0.622 \pm0.01$, $\beta \approx -11.38_{{\rm{ - }}0.24}^{{\rm{ + }}0.25}$, $\delta \approx 0.231 \pm0.004$ and $\Omega_m \approx 0.622_{{\rm{ - }}0.17}^{{\rm{ + }}0.19}$.}     
\end{figure}

\begin{figure}
	\centering
	\includegraphics[scale = 0.45]{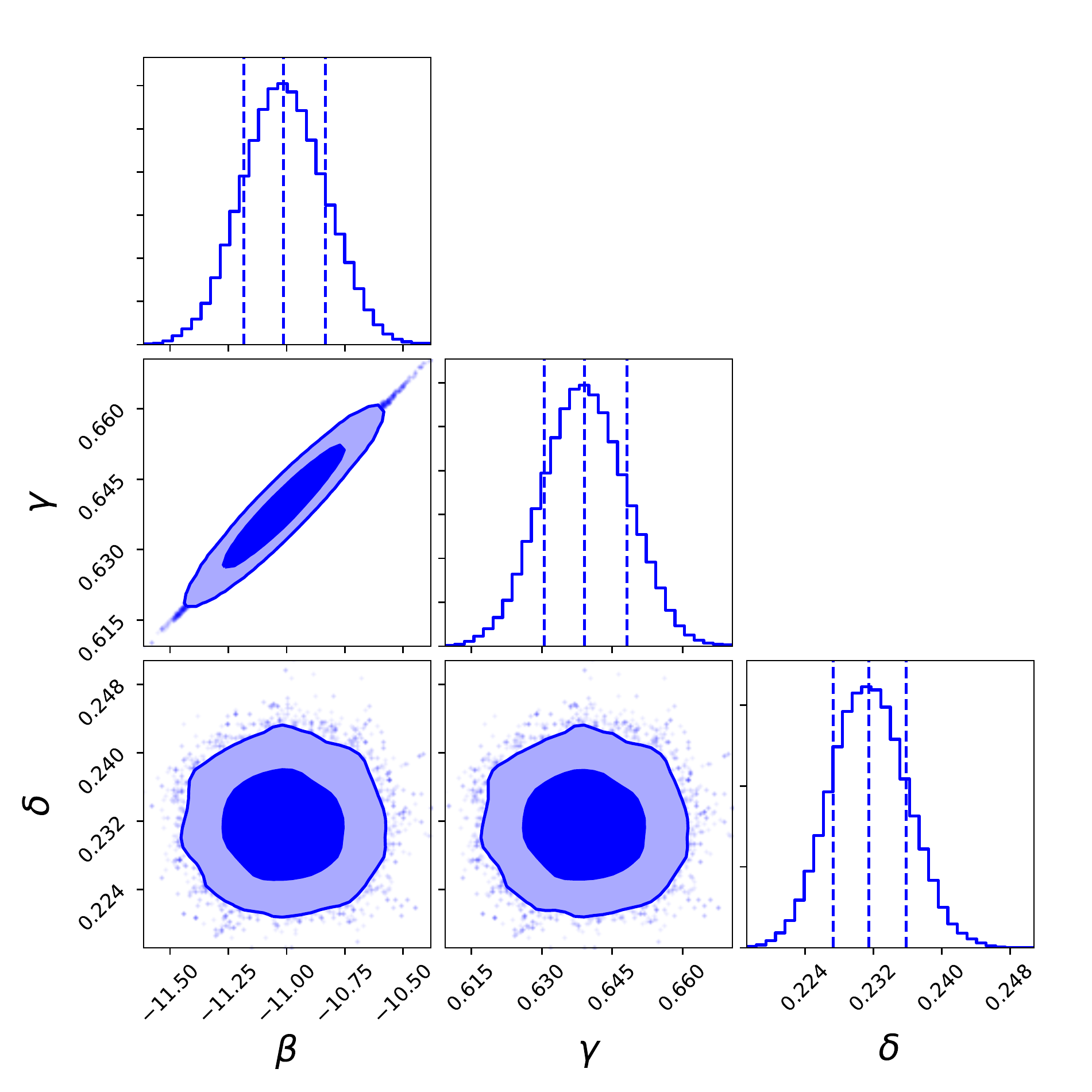}
	\label{fig2}
	\caption{Same as the FIG.1, except now for $R_{\rm  h}=ct$. The optimized values of parameters are $\gamma \approx 0.639_{{\rm{ - }}0.009}^{{\rm{ + }}0.008}$, $\beta \approx -11.01_{{\rm{ - }}0.18}^{{\rm{ + }}0.17}$ and $\delta \approx 0.231 \pm0.004$.}     
\end{figure}

\begin{figure}
	\centering
	\includegraphics[scale = 0.65]{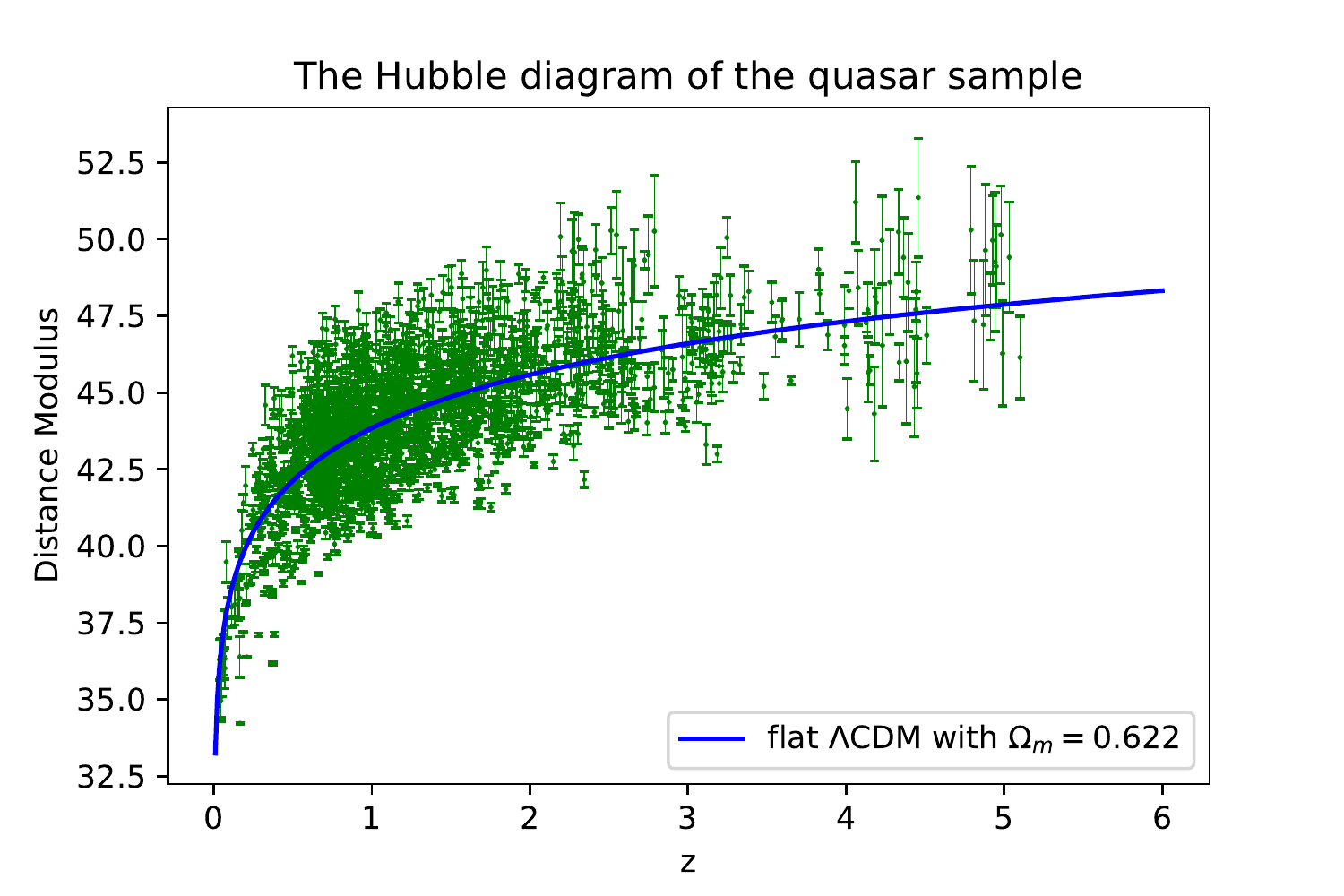}
	\label{fig3}
	\caption{The Hubble \textbf{diagram} of the quasar sample with the flat $\Lambda$CDM model for ${\Omega_m} = 0.622$, and for the best-fit curve for the sample.}     
\end{figure}

\begin{figure}
	\centering
	\includegraphics[scale = 0.65]{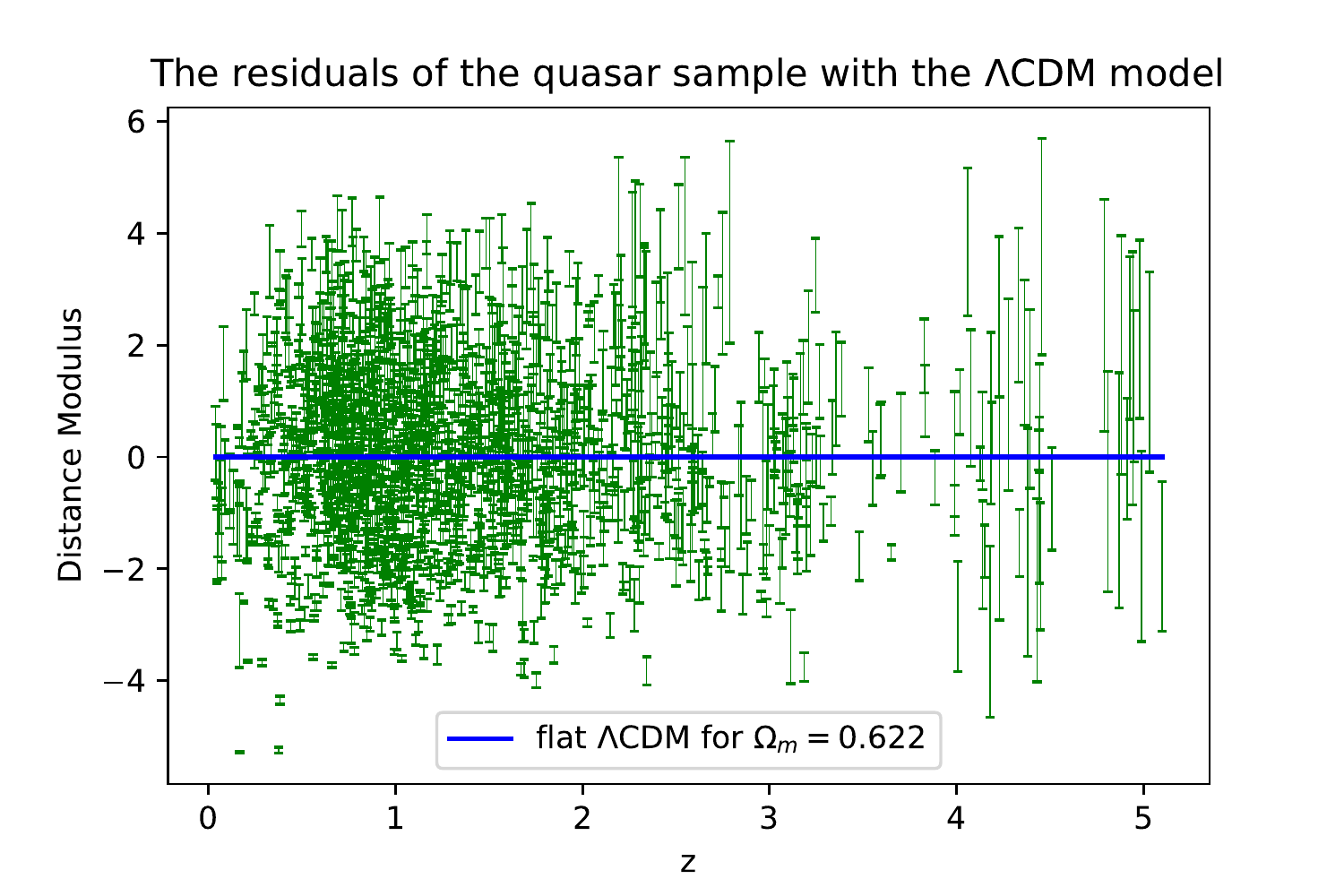}
	\label{fig4}
	\caption{The green error bars indicate the residuals of the quasar sample with  the flat $\Lambda$CDM model for ${\Omega _m} = 0.622$, and the green solid line indicates the line with zero residuals for fitting the $\Lambda$CDM model.}     
\end{figure}

\begin{figure}
	\centering
	\includegraphics[scale = 0.65]{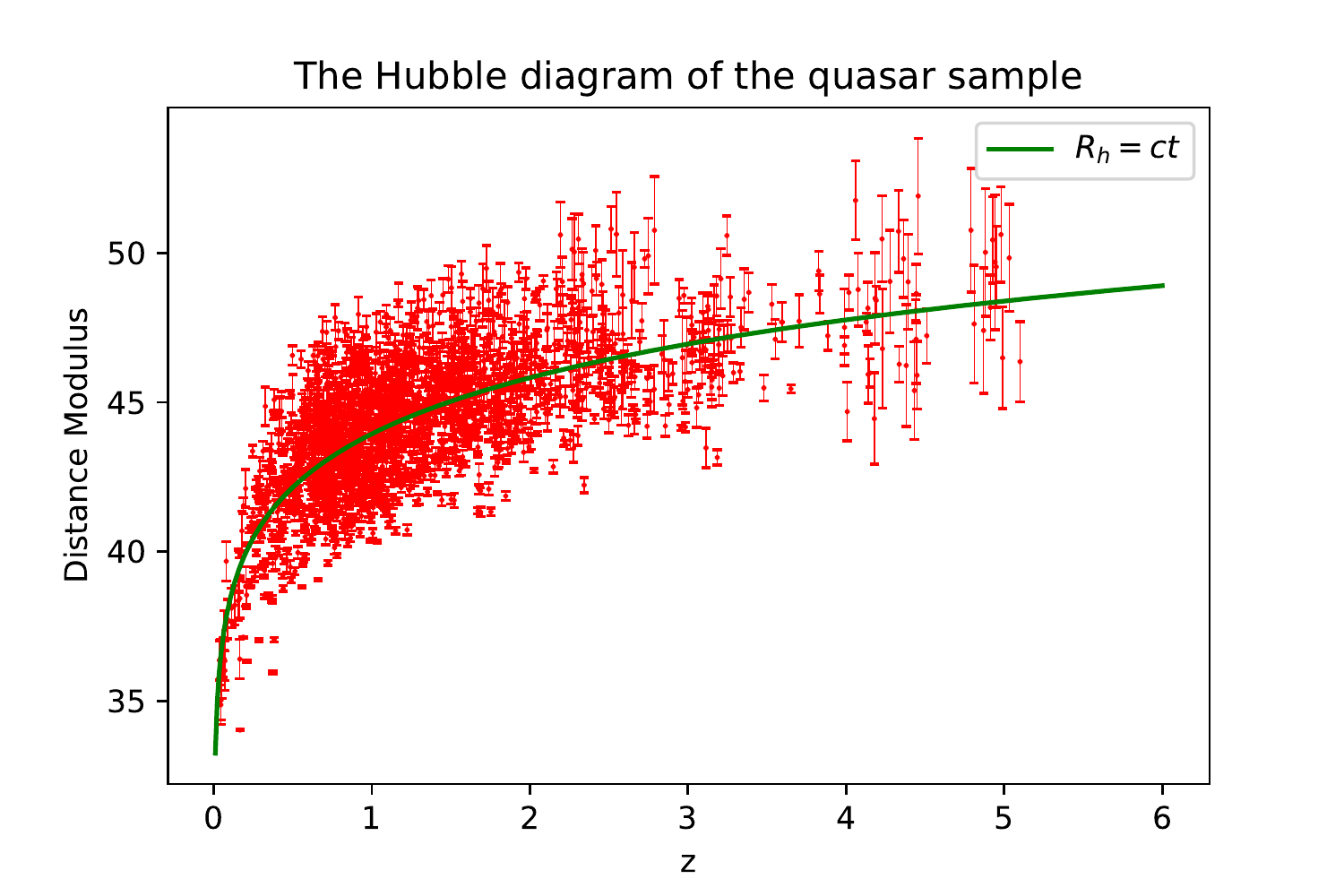}
	\label{fig5}
	\caption{The Hubble \textbf{diagram} of the quasar sample with the $R_{\rm  h}=ct$ model, 
	 and the best-fit curve for the sample.}     
\end{figure}

\begin{figure}
	\centering
	\includegraphics[scale = 0.65]{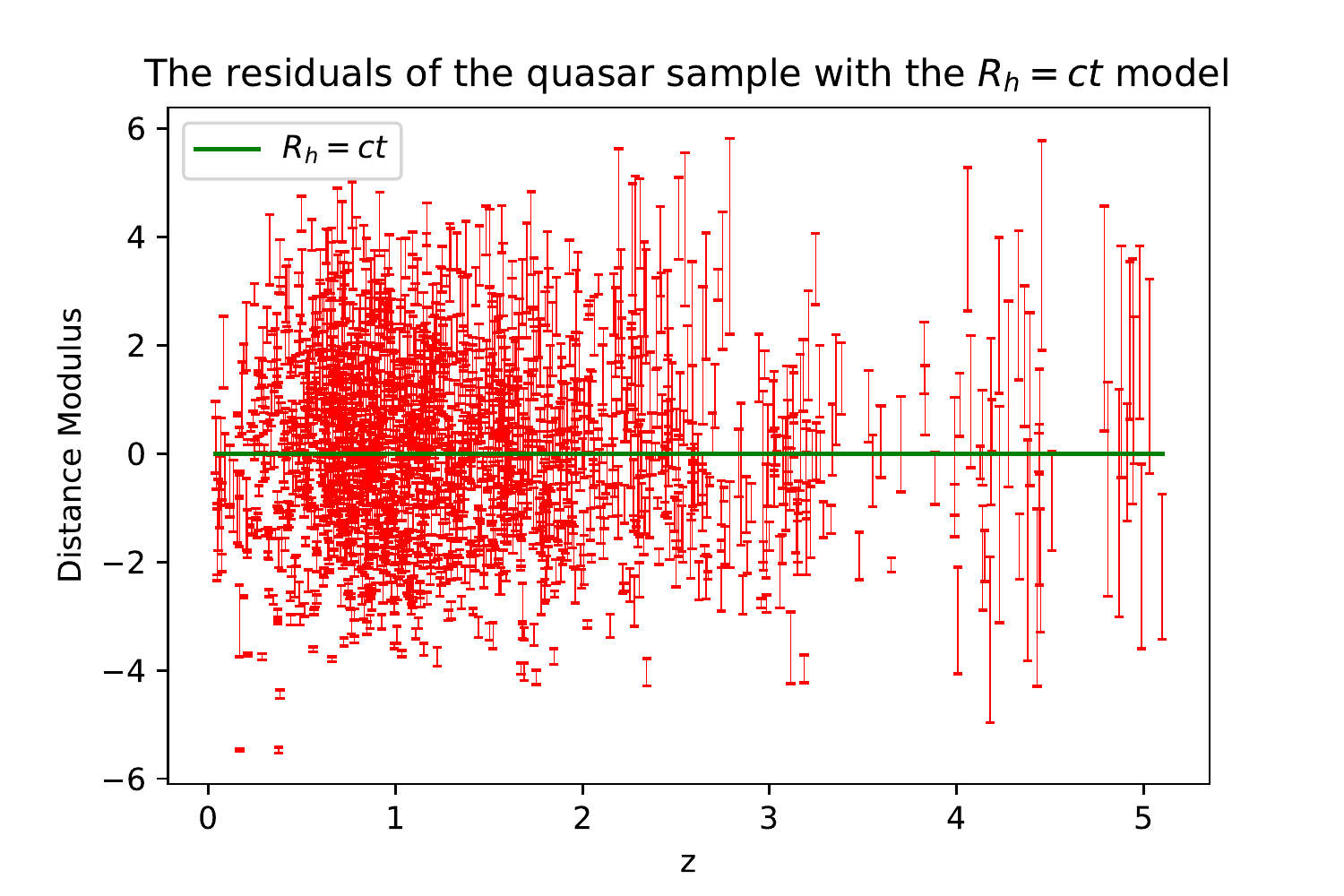}
	\label{fig6}
	\caption{The red error bars indicate the residuals of the quasar sample with the $R_{ h}=ct$ model, and the green solid line indicates the line with zero residuals for fitting the $R_{h}=ct$ model.}     
\end{figure}

In Figures (1) and (2), we present the correlation between these parameters for each model and conclude the best values for them. The light blue area and the dark blue area denote the $1\sigma$ or $2\sigma$ confidence interval respectively. As we can see, for the global intrinsic dispersion $\delta$, the optimized values for each model are highly identical. While the optimized values of $\gamma$ change slightly. The results are $\gamma \approx 0.62$ for $\Lambda$CDM and $\approx 0.64$ \textbf{for $R_{\rm  h}=ct$}. The aforementioned results are very close to the results \cite{2019NatAs...3..272R} got by using the Type Ia SNe as external calibrators. It is the high similarity of $\gamma$ and $\delta$ obtained by these two methods that further proves that the relationship assumed in Equation~(1)  can indeed be used as a standard candle for cosmological testing. Figure (3) and Figure (5) show the Hubble diagram of the quasar and give the best-fit curves for the two models in our work. There is a strange phenomenon that the value of $\Omega_{m}$ is about 0.62, which deviates from what we usually think of as around 0.31. This is the biggest difference between our work and that of ~\cite{2019MNRAS.489..517M}. The main reason is that the data is still somewhat rough, which weakens the limitation of this parameter, and the quasars at high redshifts are fitted simultaneously to obtain a relatively small distance modulus for the flat $\Lambda$CDM, which requires lower dark energy components and higher matter components. \textbf{Figure (4) and Figure (6)  show the fitted residuals using the quasar sample under the two cosmological models, respectively, and we can find that these residuals increase significantly at high redshifts ($z>1$), which may be related to the quasar sample, and the high redshift part of this sample may be of very poor quality, which affects the $\Omega_{m}$ fitted values under the $\Lambda$CDM model. However, the difference between the residuals shown on these two figures is still small.}

Since the likelihood function, Eq.~(4), which is used to optimize the parameters $\tilde \beta$, $\gamma$, and $\delta$, requires the use of luminosity distances that are different from each model, the quasar data have to be recalibrated for each model. As for the values of AIC, though the AIC of $\Lambda$CDM is smaller than $R_{\rm h}=ct$, which somewhat claims that the former is preferred, the variation is insufficient. Accordingly, the AIC likelihood plays a decisive role. The outcome is that The Akaike Information Criterion favors the $\Lambda$CDM \textbf{over} $R_{\rm h}=ct$ with a relative likelihood of $\sim 86.30\%$ versus $\sim 13.70\%$ which quantitatively testify the former is the best model based on the high-z quasars data.

It is necessary to point out that what we have done in this paper is highly similar to the work Fulvio Melia did in 2019 (\citealp*{2019MNRAS.489..517M}), but the final results between us are entirely different. Optimized parameters for each model presented in Table 1 in his paper are somewhat different from ours, particularly the value of $\Omega_{m}$ for $\Lambda$CDM \textbf{model}. They concluded that $R_{\rm h}=ct$ \textbf{model} was strongly better than $\Lambda$CDM based on BIC, but just as we have shown, the outcome is reversed based on the AIC. The main reason is precisely the variation in parameters.

\section{Conclusion}\label{sec13}

Whatever new challenges the $\Lambda$CDM model may face in the future, at least for now, we have once again validated the standard model $\Lambda$CDM based on the enhanced high-redshift quasar data. The final results demonstrate that the standard model was, and remains the best model when it confronts the constantly updated data. Another important outcome in the data process is that we affirm the reliable relationship between X-ray and UV monochromatic luminosities of quasars, as in the previous work. This relation is exactly can be a standard candle during cosmological testing. 

\section{Acknowledgments}\label{sec5}
We thank the anonymous referee for constructive comments. This work is supported by Yunnan Youth Basic Research Projects 202001AU070013.


\bibliography{sn-bibliography}


\end{document}